\documentclass{article}
\usepackage{amsfonts}
\usepackage{amsmath}

\input{tcilatex}
\begin{document}

\title{Approximate path integral solution for a Dirac particle in a deformed
Hulth\'{e}n potential}
\author{A. Kadja, F. Benamira and L. Guechi \\
Laboratoire de Physique Th\'{e}orique, D\'{e}partement de Physique, \and %
Facult\'{e} des Sciences Exactes, Universit\'{e} des Fr\`{e}res Mentouri,
Constantine, \and Route d'Ain El Bey, Constantine, Algeria}
\maketitle

\begin{abstract}
The problem of a Dirac particle moving in a deformed Hulth\'{e}n potential
is solved in the framework of the path integral formalism. With the help of
the Biedenharn transformation, the construction of a closed form for the
Green's function of the second-order Dirac equation is done by using a
proper approximation to the centrifugal term and the Green's function of the
linear Dirac equation is calculated. The energy spectrum for the bound
states is obtained from the poles of the Green's function. A Dirac particle
in the standard Hulth\'{e}n potential $\left( q=1\right) $ and a Dirac
hydrogen-like ion $\left( q=1\text{ and }a\rightarrow \infty \right) $ are
considered as particular cases.


03.65.Db-Functional analytical methods

\end{abstract}

\section{Introduction}

The Hulth\'{e}n potential has been widely used as an approximation of the
interaction potential between two bodies in a number areas in physics,
including nuclear and particle physics \cite{Bahethi}, atomic physics \cite%
{Lindhard} and molecular physics \cite{Berezin} and also in quantum
chemistry \cite{Gruninger} . It is one of the important class of
exponential-type potentials that behave like a Coulomb potential for small
values of $r$ and decrease exponentially for large values of $r$. There are
many studies in which different approaches such as the standard method based
on the Schr\"{o}dinger equation \cite{Hulthen,Flugge}, the supersymmetric
quantum mechanics technique \cite{Filho,Gonul} and the asymptotic iteration
method \cite{Bayrak} have been employed in obtaining exact or approximate
solutions of the wave equation with the Hulth\'{e}n potential for the $s-$
and $l-$waves in the framework of non-relativistic quantums mechanics.
Furthermore, a path integral treatment of this potential has been given in 
\cite{Cai}. Nevertheless, in the presence of a strong fields, the
relativistic effects on a particle under the influence of this potential
could become notable. The $s-$wave Dirac equation with a special Hulth\'{e}n
potential has been claimed exactly solved by using a constraint \cite%
{Guo,Alhaidari}. Subsequently the bound and scattering solutions of the $%
\kappa -$wave Dirac equation with the standard Hulth\'{e}n potential have
been obtained in \cite{Haouat} by adopting a suitable approximation of the
centrifugal potential term and through a technique similar to Biedenharn's 
\cite{Biedenharn}.

The purpose of the present work is to solve approximately the problem of a
relativistic particle of mass $\mu $, charge $e$ and spin $\frac{1}{2}$
moving in a deformed Hulth\'{e}n potential of the form

\begin{equation}
\text{\ }V_{q}(r)=-\frac{V_{0}}{e^{r/a}-q},\text{\ \ }  \label{a1}
\end{equation}%
where $V_{0}$ describes the depth of the potential well, $a$ is the range of
the potential and $\ q$ is a deformation parameter ($q\geq 1$). The
introduction of the parameter $q$ can serve as an additional parameter in
describing inter-atomic interactions, and especially in three-dimensional
problem, it allows to establish the center of mass location of a molecule at
a certain distance from the coordinate origin. In addition the potential $%
V_{q}(r)$ contains the standard Hulth\'{e}n potential ($q=1$) and the
Coulomb potential ($a\rightarrow \infty ,$ $q=1$) as special cases.

Here, we want to show firstly that even the problem of $s-$waves for a Dirac
particle can be solved only approximately contrary to what was claimed by
some authors \cite{Guo,Alhaidari} and secondly, as to our knowledge, there
is no path integral discussion for the relativistic Hulth\'{e}n potential,
treating the relativistic deformed Hulth\'{e}n potential (\ref{a1}) \ by
path integration, we wish to enlarge the list of problems of relativistic
particles with spin $\frac{1}{2}$ in external electric or magnetic fields 
\cite{Papadopoulos,Kayed,Boudjedaa,Bernido} which have been studied in the
framework of path integrals.

The organisation of the paper is as follows: in section 2, we first use an
approximation scheme of exponential type depending on the $q-$deformation
parameter instead the centrifugal potential term and we diagonalize the
effective Hamiltonian of the iterated form of the Dirac equation with the
help of the Biedenharn transformation \cite{Biedenharn}. Then, the Green's
function for the second order Dirac equation can be expanded into partial
waves in spherical coordinates. In section 3, the radial part of the Green's
function is converted into a path integral for the $q-$deformed Rosen-Morse
potential by using the generalized Duru-Kleinert method \cite{Kleinert}. The
procedure yields a closed form for the radial Green's function. In section
4, we construct the Green's function of the linear Dirac equation and we
obtain the energy spectrum for the $\kappa $ states. In section 5, the
Green's function and the energy spectrum associated with the standard Hulth%
\'{e}n potential and for the Coulomb potential are presented as special
cases. The section 6 will be a conclusion.

\section{Dirac equation with a deformed Hulth\'{e}n potential}

Choosing the atomic units $\hbar =c=1$, the Green's function $G(\mathbf{r}%
^{\prime \prime },\mathbf{r}^{\prime })$ for the vector potential $V_{q}(r)$
satisfies the Dirac equation

\begin{equation}
(\mu -\mathbf{M})G(\mathbf{r}^{\prime \prime },\mathbf{r}^{\prime })=\delta (%
\mathbf{r}^{\prime \prime }-\mathbf{r}^{\prime })  \label{a2}
\end{equation}%
where $\mu $ is the mass of a charged particle of spin $\frac{1}{2}$, and 
\begin{equation}
\mathbf{M}=-\mathbf{\beta \alpha .p}+\mathbf{\beta }\left( E-V_{q}(r)\right)
,  \label{a3}
\end{equation}%
in which $\mathbf{\alpha }$ and\textbf{\ }$\mathbf{\beta }$ are the usual
Dirac matrices and $E$ is the energy.

The Dirac equation (\ref{a2}) can be iterated by writing the Green's
function $G(\mathbf{r}^{\prime \prime },\mathbf{r}^{\prime })$ as

\begin{equation}
G(\mathbf{r}^{\prime \prime },\mathbf{r}^{\prime })=(\mu +\mathbf{M})g(%
\mathbf{r}^{\prime \prime },\mathbf{r}^{\prime }),  \label{a4}
\end{equation}%
where $g(\mathbf{r}^{\prime \prime },\mathbf{r}^{\prime })$ is the Green's
function defined as the solution of the second-order Dirac equation

\begin{equation}
\left( \mu ^{2}-\mathbf{M}^{2}\right) g(\mathbf{r}^{\prime \prime },\mathbf{r%
}^{\prime })=\delta (\mathbf{r}^{\prime \prime }-\mathbf{r}^{\prime }).
\label{a5}
\end{equation}%
Using the Schwinger integral representation \cite{Schwinger}, the solution
of the equation (\ref{a5}) can be written as follows:%
\begin{equation}
g(\mathbf{r}^{\prime \prime },\mathbf{r}^{\prime })=\frac{i}{2}%
\int_{0}^{\infty }d\Lambda \left\langle \mathbf{r}^{\prime \prime
}\right\vert \exp \left( -i\mathbf{H}\Lambda \right) \left\vert \mathbf{r}%
^{\prime }\right\rangle ,  \label{a6}
\end{equation}%
where the integrand $\left\langle \mathbf{r}^{\prime \prime }\right\vert
\exp \left( -i\mathbf{H}\Lambda \right) \left\vert \mathbf{r}^{\prime
}\right\rangle $ is similar to the propagator of a quantum system evolving
in $\Lambda $ time from $\mathbf{r}^{\prime }$ to $\mathbf{r}^{\prime \prime
}$ with the effective Hamiltonian%
\begin{equation}
\mathbf{H}=\frac{1}{2}\left( \mu ^{2}-\mathbf{M}^{2}\right) .  \label{a7}
\end{equation}%
Since the potential (\ref{a1}) has spherical symmetry, these are the polar
coordinates which are best suited to find the explicit expression of \ the
Green's function (\ref{a4}). Before approching the construction of the
Green's function (\ref{a6}), we note that we can make some simplifications.
Using the radial momentum operator $\mathbf{p}_{r}=\frac{1}{r}(\mathbf{r.p}%
-i)$, the Dirac operator $\mathbf{K}=\mathbf{\beta }\left( \mathbf{\sigma .L}%
+1\right) $ and the velocity operator $\mathbf{\alpha }_{r}=\frac{\mathbf{%
\alpha .r}}{r}$, the operator $\mathbf{M}$ takes the form%
\begin{equation}
\mathbf{M}=-\mathbf{\beta \alpha }_{r}\mathbf{.p}_{r}+\frac{i\mathbf{\alpha }%
_{r}\mathbf{K}}{r}+\mathbf{\beta }\left( E-V_{q}(r)\right) ,  \label{a8a}
\end{equation}%
and the effective Hamiltonian (\ref{a7}) then becomes%
\begin{equation}
\mathbf{H}=\frac{1}{2}\left( \mathbf{p}_{r}^{2}+\frac{\mathbf{K}^{2}-\mathbf{%
\beta K}}{r^{2}}-i\mathbf{\alpha }_{r}\frac{d}{dr}V_{q}(r)-\left(
E-V_{q}(r)\right) ^{2}+\mu ^{2}\right) .  \label{a8b}
\end{equation}%
Note that $\mathbf{\beta }$ has eigenvalues $\widetilde{\beta }=\pm 1$ and $%
\mathbf{K}^{2}=\mathbf{J}^{2}+\frac{1}{4}$, where $\mathbf{J}=\mathbf{L}+%
\frac{1}{2}\mathbf{\sigma }$ is the total angular momentum operator. The
eigenvalues of the operator $\mathbf{J}^{2}$ are $j(j+1),$ where $j=l\pm 
\frac{1}{2}$, except for the s-states $(l=0)$, in which $j$ can take only
the value $\frac{1}{2}$. Thus, the operator $\mathbf{K}^{2}$ has eigenvalues 
$\kappa ^{2}=\left( j+\frac{1}{2}\right) ^{2}$, so that $\kappa =\pm \left(
j+\frac{1}{2}\right) =\pm 1,\pm 2,\pm 3,...$ .

On the other hand, the Hulth\'{e}n-type potential in (\ref{a8b}) depends on
the arbitrary real deformation parameter $q$ and the Green's function (\ref%
{a6}) cannot be evaluated exactly because of the presence of the centrifugal
potential term. However, if $r/a\ll 1$, it is easy to show that $\frac{%
qe^{r/a}}{a^{2}\left( e^{r/a}-q\right) ^{2}}+\frac{1}{12a^{2}}$ can be used
as a good approximation to $\frac{1}{r^{2}}$ in the centrifugal term when
the parameter $q\geq 1.$Then, by means of this approximation, the effective
Hamiltonian (\ref{a8b}) can be written in the form

\begin{equation}
\mathbf{H}=\frac{1}{2}\left( \mathbf{p}_{r}^{2}+\mathbf{\Gamma }(\mathbf{%
\Gamma }+1)\frac{qe^{r/a}}{a^{2}\left( e^{r/a}-q\right) ^{2}}+\frac{%
V_{0}\left( \frac{V_{0}}{q}-2E\right) }{e^{r/a}-q}+\frac{1}{12a^{2}}\mathbf{K%
}\left( \mathbf{K}-\mathbf{\beta }\right) +\mu ^{2}-E^{2}\right) ,
\label{a9}
\end{equation}%
where

\begin{equation}
\mathbf{\Gamma }=-\left( \mathbf{\beta K}+i\frac{aV_{0}}{q}\mathbf{\alpha }%
_{r}\right)  \label{a10}
\end{equation}%
is the Martin-Glauber operator \cite{Martin,Wong}. In analogy with the work
of Biedenharn on the Coulomb problem, we have found a similarity
transformation $\mathbf{S}$ defined by%
\begin{equation}
\mathbf{S}=\exp \left[ \frac{i}{2}\mathbf{\beta \alpha }_{r}\tanh
^{-1}\left( \frac{aV_{0}}{q\mathbf{K}}\right) \right]  \label{a11}
\end{equation}%
which diagonalizes $\Gamma .$ This gives%
\begin{equation}
\Gamma _{s}=\mathbf{S}\Gamma \mathbf{S}^{-1}=-\mathbf{\beta K}\left[
1-\left( \frac{aV_{0}}{q\mathbf{K}}\right) ^{2}\right] ^{\frac{1}{2}},
\label{a12}
\end{equation}%
whose eigenvalues are, in view of those of $\mathbf{\beta }$ and $\mathbf{K}%
^{2}$:%
\begin{equation}
\gamma =\pm \left[ \kappa ^{2}-\left( \frac{aV_{0}}{q}\right) ^{2}\right] ^{%
\frac{1}{2}}.  \label{a13a}
\end{equation}

Under the transformation (\ref{a11}), equation (\ref{a4}) is thus put in the
form:%
\begin{equation}
\widetilde{G}(\mathbf{r}^{\prime \prime },\mathbf{r}^{\prime })=(\mu +%
\mathbf{M}_{s})\widetilde{g}(\mathbf{r}^{\prime \prime },\mathbf{r}^{\prime
}),  \label{a13b}
\end{equation}%
where%
\begin{equation}
\mathbf{M}_{s}=\mathbf{SMS}^{-1},  \label{a13c}
\end{equation}%
and 
\begin{equation}
\widetilde{g}(\mathbf{r}^{\prime \prime },\mathbf{r}^{\prime })=\frac{i}{2}%
\int_{0}^{\infty }d\Lambda \left\langle \mathbf{r}^{\prime \prime
}\right\vert \exp \left( -i\mathbf{H}_{S}\Lambda \right) \left\vert \mathbf{r%
}^{\prime }\right\rangle ,  \label{a13d}
\end{equation}%
with 
\begin{equation}
\mathbf{H}_{s}=\mathbf{SHS}^{-1}.  \label{a13e}
\end{equation}

The integrand in Eq. (\ref{a13d}) can be expressed into partial wave
expansion as follows:%
\begin{equation}
\left\langle \mathbf{r}^{\prime \prime }\right\vert \exp \left( -i\mathbf{H}%
_{s}\Lambda \right) \left\vert \mathbf{r}^{\prime }\right\rangle =\frac{1}{%
r^{\prime \prime }r^{\prime }}\underset{\lambda }{\sum }\left\langle \theta
^{\prime \prime },\phi ^{\prime \prime }\right\vert \left. \lambda
\right\rangle \left\langle r^{\prime \prime }\right\vert \exp \left( -i%
\mathbf{H}_{\lambda }\Lambda \right) \left\vert r^{\prime }\right\rangle
\left\langle \lambda \right\vert \left. \theta ^{\prime },\varphi ^{\prime
}\right\rangle ,\varphi  \label{a14}
\end{equation}%
where we have denoted by $\left\vert \lambda \right\rangle =\left\vert
j,m,\kappa ;\widetilde{\beta }\right\rangle $ the simultaneous eigenvectors
of the operators $\mathbf{J}^{2}$, $\mathbf{J}_{z}$, $\mathbf{K}$, $\mathbf{%
\beta }$, and by $\left\langle r^{\prime \prime }\right\vert \exp \left( -i%
\mathbf{H}_{\lambda }\Lambda \right) \left\vert r^{\prime }\right\rangle $
the radial propagator, in which%
\begin{equation}
\mathbf{H}_{\lambda }=\frac{1}{2}\left( \mathbf{p}_{r}^{2}+\lambda (\lambda
+1)\frac{qe^{r/a}}{a^{2}\left( e^{r/a}-q\right) ^{2}}+\frac{V_{0}\left( 
\frac{V_{0}}{q}-2E\right) }{e^{r/a}-q}+\mu ^{2}-E^{2}+\frac{\kappa \left(
\kappa -\widetilde{\beta }\right) }{12a^{2}}\right) ,  \label{a15}
\end{equation}%
with%
\begin{equation}
\lambda =\left\vert \gamma \right\vert +\frac{1}{2}\left( \text{sign }\gamma
-1\right) .\text{ \ }1/(r/a)^{2}  \label{a16}
\end{equation}%
and by $\left\langle \theta ,\phi \right\vert \left. j,m,\kappa ,\widetilde{%
\beta }\right\rangle $ the dependence in terms of angles $\theta ,\phi ,$
spin and $\widetilde{\beta }$ variables which can be written explicitly as:%
\begin{equation}
\left\{ 
\begin{array}{c}
\left\langle \theta ,\phi \right\vert \left. j,m,\kappa ,-1\right\rangle
=\left( 
\begin{array}{l}
0 \\ 
\Omega _{\kappa }^{m}\left( \theta ,\phi \right)%
\end{array}%
\right) , \\ 
\left\langle \theta ,\phi \right\vert \left. j,m,\kappa ,1\right\rangle
=\left( 
\begin{array}{l}
\Omega _{-\kappa }^{m}\left( \theta ,\phi \right) \\ 
0%
\end{array}%
\right) ,%
\end{array}%
\right.  \label{a17}
\end{equation}%
where $\Omega _{k}^{m}\left( \theta ,\phi \right) $ denotes the
hyperspherical harmonic%
\begin{eqnarray}
\Omega _{\kappa }^{m}\left( \theta ,\phi \right) &=&-\text{sgn }\kappa \sqrt{%
\frac{\kappa -m+\frac{1}{2}}{2\kappa +1}}\chi _{\frac{1}{2}}^{\frac{1}{2}%
}Y_{l}^{m-\frac{1}{2}}\left( \theta ,\phi \right) +\sqrt{\frac{\kappa +m+%
\frac{1}{2}}{2\kappa +1}}\chi _{\frac{1}{2}}^{-\frac{1}{2}}Y_{l}^{m+\frac{1}{%
2}}\left( \theta ,\phi \right)  \notag \\
&&  \label{a18}
\end{eqnarray}%
in which $\chi _{\frac{1}{2}}^{\frac{1}{2}}$ and $\chi _{\frac{1}{2}}^{-%
\frac{1}{2}}$ are the spin wavefunctions, and sgn $\kappa =\pm $sgn $\gamma $
for $\widetilde{\beta }=\mp 1.$

Using (\ref{a17}), the propagator (\ref{a14}) can be written as:%
\begin{equation}
\left\langle \mathbf{r}^{\prime \prime }\right\vert \exp \left( -i\mathbf{H}%
_{s}\Lambda \right) \left\vert \mathbf{r}^{\prime }\right\rangle =\frac{1}{%
r^{\prime }r^{\prime \prime }}\underset{j,\kappa }{\sum }\left\langle
r^{\prime \prime }\right\vert \exp \left( -i\mathbf{H}_{\lambda }\Lambda
\right) \left\vert r^{\prime }\right\rangle \Omega _{\kappa ,\kappa
}^{j}\left( \theta ^{\prime \prime },\phi ^{\prime \prime },\theta ^{\prime
},\phi ^{\prime }\right) \beta ^{2},  \label{a19}
\end{equation}%
where%
\begin{equation}
\Omega _{\kappa ,\kappa }^{j}\left( \theta ^{\prime \prime },\phi ^{\prime
\prime },\theta ^{\prime },\phi ^{\prime }\right) =\underset{m}{\sum }\Omega
_{\kappa }^{m}\left( \theta ^{\prime \prime },\phi ^{\prime \prime }\right)
\Omega _{\kappa }^{m\ast }\left( \theta ^{\prime },\phi ^{\prime }\right) .
\label{a20}
\end{equation}

Inserting (\ref{a19}) into (\ref{a13d}), we obtain

\begin{equation}
\widetilde{g}(\mathbf{r}^{\prime \prime },\mathbf{r}^{\prime })=\frac{1}{%
r^{\prime \prime }r^{\prime }}\underset{j,\kappa }{\sum }g_{j,\kappa }\left(
r^{\prime \prime },r^{\prime }\right) \Omega _{\kappa ,\kappa }^{j}\left(
\theta ^{\prime \prime },\phi ^{\prime \prime },\theta ^{\prime },\phi
^{\prime }\right) \beta ^{2}  \label{a21}
\end{equation}%
with the radial Green's function given by%
\begin{eqnarray}
g_{j,\kappa }\left( r^{\prime \prime },r^{\prime }\right) &=&\frac{i}{2}%
\int_{0}^{\infty }d\Lambda \left\langle r^{\prime \prime }\right\vert \exp
\left\{ -\frac{i}{2}\left[ \mathbf{p}_{r}^{2}+\frac{q\lambda \left( \lambda
+1\right) e^{r/a}}{a^{2}\left( e^{r/a}-q\right) ^{2}}\right. \right.  \notag
\\
&&\left. \left. +\frac{V_{0}\left( \frac{V_{0}}{q}-2E\right) }{e^{r/a}-q}%
+\epsilon ^{2}\right] \Lambda \right\} \left\vert r^{\prime }\right\rangle
\label{a22}
\end{eqnarray}%
where%
\begin{equation}
\epsilon =\sqrt{\mu ^{2}-E^{2}+\frac{\kappa (\kappa -\widetilde{\beta })}{%
12a^{2}}}.  \label{a23}
\end{equation}%
The expression (\ref{a22}) can now be solved by the path integral technique.

\section{Path integral solution}

It is easy to express (\ref{a22}) in the form of a path integral, 
\begin{equation}
g_{j,\kappa }\left( r^{\prime \prime },r^{\prime }\right) =\frac{i}{2}%
\int_{0}^{\infty }dS^{\prime }P_{j,\kappa }\left( r^{\prime \prime
},r^{\prime };S^{\prime }\right) ,  \label{a24}
\end{equation}%
where 
\begin{eqnarray}
P_{j,\kappa }\left( r^{\prime \prime },r^{\prime };S^{\prime }\right)
&=&f_{R}(r^{\prime \prime })f_{L}(r^{\prime })\left\langle r^{\prime \prime
}\right\vert \exp \left\{ -\frac{i}{2}f_{L}(r)\left[ \mathbf{p}_{r}^{2}+%
\frac{q\lambda \left( \lambda +1\right) e^{r/a}}{a^{2}\left(
e^{r/a}-q\right) ^{2}}\right. \right.  \notag \\
&&\left. \left. +\frac{V_{0}\left( \frac{V_{0}}{q}-2E\right) }{e^{r/a}-q}%
+\epsilon ^{2}\right] f_{R}(r)S^{\prime }\right\} \left\vert r^{\prime
}\right\rangle  \notag \\
&=&f_{R}(r^{\prime \prime })f_{L}(r^{\prime })\underset{N\rightarrow \infty }%
{\lim }\overset{N}{\underset{n=1}{\dprod }}\left[ \int dr_{n}\right] \overset%
{N+1}{\underset{n=1}{\dprod }}\left[ \int \frac{d\left( p_{r}\right) _{n}}{%
2\pi }\right]  \notag \\
&&\times \exp \left\{ i\overset{N+1}{\underset{n=1}{\sum }}\mathcal{A}%
_{1}^{n}\right\} .  \label{a25}
\end{eqnarray}%
Here $f_{L}(r)$ and $f_{R}(r)$ \ are the regulating functions defined by
Kleinert \cite{Kleinert} as:%
\begin{equation}
f(r)=f_{L}(r)f_{R}(r)=f^{1-\delta }(r)f^{\delta }(r),  \label{a26}
\end{equation}%
where $\delta $ is the splitting parameter and $\mathcal{A}_{1}^{n}$, the
modified short-time action in the canonical form. With the notation, $%
\varepsilon _{s^{\prime }}=\frac{S^{\prime }}{N+1}=ds^{\prime }=\frac{ds}{%
f_{L}(r_{n})f_{R}(r_{n-1})},$ $ds=\varepsilon _{s}=\frac{\Lambda }{N+1},$ $%
\left( p_{r}\right) _{n}=p_{r}(s_{n}),$ $r_{n}=r(s_{n})$ and $\Delta
r_{n}=r_{n}-r_{n-1},$ we have 
\begin{eqnarray}
\mathcal{A}_{1}^{n} &=&-\left( p_{r}\right) _{n}\Delta r_{n}-\frac{%
\varepsilon _{s^{\prime }}}{2}f_{L}\left( r_{n}\right) \left[ \left(
p_{r}\right) _{n}^{2}+\frac{q\lambda \left( \lambda +1\right) e^{r_{n}/a}}{%
a^{2}\left( e^{r_{n}/a}-q\right) ^{2}}\right.  \notag \\
&&\left. +\frac{V_{0}\left( \frac{V_{0}}{q}-2E\right) }{e^{r_{n}/a}-q}%
+\epsilon ^{2}\right] f_{R}(r_{n-1}).\text{ \ \ \ }\widetilde{\varepsilon }
\label{a27}
\end{eqnarray}

In order to simplify the calculation of the kernel $P_{j,k}\left( r^{\prime
\prime },r^{\prime };S^{\prime }\right) $, let us put the splitting
parameter $\delta =\frac{1}{2}$, that is to say, we choose the mid-point
prescription. This can be justified by the fact that the final result is
independent of this parameter. Then, by integrating with respect to the
momentum variables $\left( p_{r}\right) _{n}$, we find that%
\begin{eqnarray}
P_{j,\kappa }\left( r^{\prime \prime },r^{\prime };S^{\prime }\right) &=& 
\left[ f^{\prime }\left( r^{\prime \prime }\right) f(r^{\prime })\right] ^{%
\frac{1}{4}}\underset{N\rightarrow \infty }{\lim }\overset{N+1}{\underset{n=1%
}{\dprod }}\left[ \frac{1}{2i\pi \varepsilon _{s^{\prime }}}\right] ^{\frac{1%
}{2}}  \notag \\
&&\times \overset{N}{\underset{n=1}{\dprod }}\left[ \int \frac{dr_{n}}{\sqrt{%
f(r_{n})}}\right] \exp \left\{ i\overset{N+1}{\underset{n=1}{\sum }}\mathcal{%
A}_{2}^{n}\right\} ,  \label{a28}
\end{eqnarray}%
with the short-time action in configuration space given by%
\begin{eqnarray}
\mathcal{A}_{2}^{n} &=&\frac{\left( \Delta r_{n}\right) ^{2}}{2\varepsilon
_{s^{\prime }}\sqrt{f(r_{n})f(r_{n-1})}}-\frac{\varepsilon _{s^{\prime }}}{2}%
\left[ \frac{q\lambda \left( \lambda +1\right) e^{r_{n}/a}}{a^{2}\left(
e^{r_{n}/a}-q\right) ^{2}}\right.  \notag \\
&&\left. +\frac{V_{0}\left( \frac{V_{0}}{q}-2E\right) }{e^{r_{n}/a}-q}%
+\epsilon ^{2}\right] \sqrt{f(r_{n})f(r_{n-1})}.  \label{a29}
\end{eqnarray}

To evaluate the path integral (\ref{a28}), we perform the following space
transformation:%
\begin{equation}
r\in \left] r_{0};+\infty \right[ \rightarrow \xi \in \left] -\infty
;+\infty \right[  \label{a30}
\end{equation}%
defined by%
\begin{equation}
r=a\ln \left[ \exp \left( 2\xi /a\right) +q\right] ,  \label{a31}
\end{equation}%
accompanied by the appropriate regulating function

\begin{equation}
f(r(\xi ))=\frac{e^{2\xi /a}}{\cosh _{q}^{2}\left( \xi /a\right) }=g^{\prime
2}\left( \xi \right) .  \label{a32}
\end{equation}

Under these transformations, the kernel (\ref{a28}) takes the form:

\begin{eqnarray}
P_{j,\kappa }\left( r^{\prime \prime },r^{\prime };S^{\prime }\right) &=& 
\left[ f^{\prime }\left( r^{\prime \prime }\right) f(r^{\prime })\right] ^{%
\frac{1}{4}}\underset{N\rightarrow \infty }{\lim }\overset{N+1}{\underset{n=1%
}{\dprod }}\left[ \frac{1}{2i\pi \varepsilon _{s^{\prime }}}\right] ^{\frac{1%
}{2}}\overset{N}{\underset{n=1}{\dprod }}\left[ \int d\xi _{n}\right] \text{
\ \ \ \ \ \ \ }  \notag \\
&&\times \exp \left\{ i\overset{N+1}{\underset{n=1}{\sum }}\left[ \frac{%
\left( \Delta \xi _{n}\right) ^{2}}{2\varepsilon _{s^{\prime }}}+\frac{1}{%
8\varepsilon _{s^{\prime }}}\left( \left( \frac{g^{\prime \prime }}{%
g^{\prime }}\right) ^{2}-\frac{2}{3}\frac{g^{\prime \prime \prime }}{%
g^{\prime }}\right) \left( \Delta \xi _{n}\right) ^{4}\right. \right.  \notag
\\
&&-\frac{\varepsilon _{s^{\prime }}}{a^{2}}\left[ \widetilde{\epsilon }%
^{2}+\lambda \left( \lambda +1\right) \right] +\frac{\varepsilon _{s^{\prime
}}}{2a^{2}}\frac{\left( q\widetilde{\epsilon }^{2}-\omega ^{2}\right) }{%
\cosh _{q}^{2}\left( \xi _{n}/a\right) }  \notag \\
&&\left. \left. -\frac{\varepsilon _{s^{\prime }}}{a^{2}}\left[ \widetilde{%
\epsilon }^{2}-\lambda \left( \lambda +1\right) \right] \tanh _{q}\left( \xi
_{n}/a\right) \right] \right\} ,\text{ \ \ \ \ \ \ }  \label{a33}
\end{eqnarray}%
where%
\begin{equation}
\widetilde{\epsilon }=a\sqrt{\mu ^{2}-E^{2}+\frac{\kappa (\kappa -\widetilde{%
\beta })}{12a^{2}}},  \label{a34a}
\end{equation}%
and%
\begin{equation}
\omega =a\sqrt{V_{0}\left( \frac{V_{0}}{q}-2E\right) }.  \label{a34b}
\end{equation}%
In Eqs. (\ref{a32}) and (\ref{a33}), we have used the deformed hyperbolic
functions introduced for the first time by Arai \cite{Arai} and denoted by 
\begin{equation}
\cosh _{q}x=\frac{1}{2}\left( e^{x}+qe^{-x}\right) ,\text{ \ }\sinh _{q}x=%
\frac{1}{2}\left( e^{x}-qe^{-x}\right) ,\text{ \ \ \ tanh}_{q}x=\frac{\sinh
_{q}x}{\cosh _{q}x},  \label{a35}
\end{equation}%
where q is a real parameter.

Note that the term in $\left( \Delta \xi _{n}\right) ^{4}$ appearing in the
action contained in Eq. (\ref{a33}) contributes significantly to the path
integral. It can be estimated by using the Mc Laughlin and Schulman
procedure \cite{Schulman} and replaced by

\begin{equation}
\left\langle (\Delta \xi _{n})^{4}\right\rangle =\int_{-\infty }^{\infty
}d(\Delta \xi _{n})(\Delta \xi _{n})^{4}\left( \frac{1}{2i\pi \varepsilon
_{s^{\prime }}}\right) ^{\frac{1}{2}}\exp \left[ \frac{i}{2\varepsilon
_{s^{\prime }}}(\Delta \xi _{n})^{2}\right] =-3\varepsilon _{s^{\prime
}}^{2}.  \label{a36}
\end{equation}

After changing $\alpha \xi _{n}$ into $y_{n}$ and $\varepsilon _{s^{\prime
}} $ into $\alpha ^{-2}\varepsilon _{\tau }$, one obtains for the radial
Green's function (\ref{a24}) the following expression:

\begin{equation}
g_{j,\kappa }\left( r^{\prime \prime },r^{\prime }\right) =\frac{a}{2}\left[
f\left( r^{\prime \prime }\right) f(r^{\prime })\right] ^{\frac{1}{4}%
}G_{RM}(y^{\prime \prime },y^{\prime };\widetilde{E}),  \label{a37}
\end{equation}%
where%
\begin{equation}
G_{RM}(y^{\prime \prime },y^{\prime };\widetilde{E})=i\int_{0}^{\infty
}d\sigma \exp (i\widetilde{E}\sigma )P_{j,\kappa }^{RM}(y^{\prime \prime
},y^{\prime };\sigma ),  \label{a38}
\end{equation}%
with%
\begin{equation}
\widetilde{E}=-\left( \widetilde{\epsilon }^{2}+\lambda \left( \lambda
+1\right) +\frac{1}{4}\right)  \label{a39}
\end{equation}%
and%
\begin{equation}
P_{j,\kappa }^{RM}(y^{\prime \prime },y^{\prime };\sigma )=\int \mathcal{D}%
y(\tau )\exp \left\{ i\int_{0}^{\sigma }\left[ \frac{\overset{.}{y}^{2}}{2}%
-V_{j,\kappa }^{RM}(y)\right] d\tau \right\}  \label{a40}
\end{equation}%
is the propagator for the general Rosen-Morse potential \cite{Rosen} defined
in terms of $q$-deformed hyperbolic functions as:%
\begin{equation}
V_{j,\kappa }^{RM}(y)=A\tanh _{q}y-\frac{B}{\cosh _{q}^{2}y};\text{ \ \ \ \
\ }y\in 
\mathbb{R}
,  \label{a41}
\end{equation}%
where the constants $A$ and $B$ are given by%
\begin{equation}
\left\{ 
\begin{array}{c}
A=\widetilde{\epsilon }^{2}-\lambda \left( \lambda +1\right) -\frac{1}{4},
\\ 
B=\frac{1}{2}\left( q\widetilde{\epsilon }^{2}-\omega ^{2}-\frac{q}{4}%
\right) .%
\end{array}%
\right.  \label{a42}
\end{equation}%
Since its path integral solution is well-established \cite%
{Grosche,Benamira1,Benamira2} , we immediatly can write down the closed
expression for the Green's function as:

\begin{eqnarray}
G_{RM}(y^{\prime \prime },y^{\prime };\widetilde{E}) &=&\frac{\Gamma \left(
M_{1}-L_{\widetilde{E}}\right) \Gamma \left( L_{\widetilde{E}%
}+M_{1}+1\right) }{\Gamma \left( M_{1}+M_{2}+1\right) \Gamma (M_{1}-M_{2}+1)}
\notag \\
&&\times \left( \frac{1-\tanh _{q}y^{\prime }}{2}\frac{1-\tanh
_{q}y^{^{\prime \prime }}}{2}\right) ^{\frac{M_{1}+M_{2}}{2}}  \notag \\
&&\times \left( \frac{1+\tanh _{q}y^{\prime }}{2}\frac{1+\tanh
_{q}y^{^{\prime \prime }}}{2}\right) ^{\frac{M_{1}-M_{2}}{2}}  \notag \\
&&\times \text{ }_{2}F_{1}\left( M_{1}-L_{\widetilde{E}},L_{\widetilde{E}%
}+M_{1}+1,M_{1}+M_{2}+1;\frac{1-\tanh _{q}y_{<}}{2}\right)  \notag \\
&&\times \text{ }_{2}F_{1}\left( M_{1}-L_{\widetilde{E}},L_{\widetilde{E}%
}+M_{1}+1,M_{1}-M_{2}+1;\frac{1+\tanh _{q}y_{>}}{2}\right) ,  \notag \\
&&  \label{a43}
\end{eqnarray}%
where we have used the notation%
\begin{equation}
\left\{ 
\begin{array}{c}
L_{\widetilde{E}}=-\frac{1}{2}+\left( \frac{1}{16}+2E_{PT^{\prime }}\right)
^{\frac{1}{2}} \\ 
E_{PT^{\prime }}=\frac{1}{2}\left( \widetilde{\epsilon }^{2}-\frac{\omega
^{2}}{q}-\frac{1}{16}\right) , \\ 
M_{1,2}=\widetilde{\epsilon }\pm \left( \lambda +\frac{1}{2}\right) ,%
\end{array}%
\right.  \label{a44}
\end{equation}%
$_{2}F_{1}(a,b,c;z)$ is the hypergeometric function and the symbols $y_{>}$
and $y_{<}$ denote max$(y^{\prime \prime },y^{\prime })$ and min$(y^{\prime
\prime },y^{\prime })$, respectively.

\section{Green's function and energy spectrum}

Having completed the path integration, we substitute (\ref{a43}) into (\ref%
{a37}) and then into (\ref{a21}). Then, after transforming the variable $%
y=\xi /a$ to the radial variable by (\ref{a31}) and taking into account the
notation (\ref{a44}), we get the following expression for the complete
Green's function of the second order Dirac equation:%
\begin{eqnarray}
\widetilde{g}(\mathbf{r}^{\prime \prime },\mathbf{r}^{\prime }) &=&\frac{a}{%
r^{\prime \prime }r^{\prime }}\underset{j,\kappa }{\sum }\frac{\Gamma \left(
1+\lambda +\widetilde{\epsilon }-\sqrt{\widetilde{\epsilon }^{2}-\frac{%
\omega ^{2}}{q}}\right) \Gamma \left( 1+\lambda +\widetilde{\epsilon }+\sqrt{%
\widetilde{\epsilon }^{2}-\frac{\omega ^{2}}{q}}\right) }{\Gamma \left( 2%
\widetilde{\epsilon }+1\right) \Gamma (2\lambda +2)}  \notag \\
&&\times \left( q^{2}e^{-(r^{\prime \prime }+r^{\prime })/a}\right) ^{%
\widetilde{\epsilon }}\left[ \left( 1-qe^{-r^{\prime \prime }/a}\right)
\left( 1-qe^{-r^{\prime }/a}\right) \right] ^{\lambda +1}  \notag \\
&&\times \text{ }_{2}F_{1}\left( 1+\lambda +\widetilde{\epsilon }-\sqrt{%
\widetilde{\epsilon }^{2}-\frac{\omega ^{2}}{q}},1+\lambda +\widetilde{%
\epsilon }+\sqrt{\widetilde{\epsilon }^{2}-\frac{\omega ^{2}}{q}},2%
\widetilde{\epsilon }+1;qe^{-r^{\prime \prime }/a}\right)  \notag \\
&&\times \text{ }_{2}F_{1}\left( 1+\lambda +\widetilde{\epsilon }-\sqrt{%
\widetilde{\epsilon }^{2}-\frac{\omega ^{2}}{q}},1+\lambda +\widetilde{%
\epsilon }+\sqrt{\widetilde{\epsilon }^{2}-\frac{\omega ^{2}}{q}},2\lambda
+2;1-qe^{-r^{\prime }/a}\right)  \notag \\
&&\times \Omega _{\kappa ,\kappa }^{j}\left( \theta ^{\prime \prime },\phi
^{\prime \prime },\theta ^{\prime },\phi ^{\prime }\right) \beta ^{2}.
\label{a45}
\end{eqnarray}

We now proceed to evaluate the Green's function for the linear Dirac
equation (\ref{a13b}). To do this, we first note that the transformed
operator $\mathbf{M}_{s}$ applied to $\kappa -$states can be put in the form%
\begin{equation}
\mathbf{M}_{s}=i\mathbf{\beta }\alpha _{r}\left[ \frac{d}{dr}+\frac{1-\gamma
\beta }{r}+\frac{aV_{0}E}{q\gamma }\beta \right] +\frac{\kappa E}{\gamma }%
\beta \mathbf{+}\frac{V_{0}\beta }{e^{r/a}-q}-\frac{aV_{0}\beta }{qr}.
\label{a46}
\end{equation}%
Then, on account of the relations%
\begin{equation}
\left\{ 
\begin{array}{c}
\alpha _{r}=\sigma _{r}\gamma ^{5}=i\sigma _{r}\beta \gamma ^{1}\gamma
^{2}\gamma ^{3}, \\ 
\gamma ^{i}=\beta \alpha _{i};\text{ }i=1,2,3,%
\end{array}%
\right.  \label{a47}
\end{equation}%
and the equation%
\begin{equation}
\sigma _{r}\Omega _{\kappa }^{m}\left( \theta ,\phi \right) =-\Omega
_{-\kappa }^{m}\left( \theta ,\phi \right) ,  \label{a48}
\end{equation}%
we arrive at the following expression:

\begin{eqnarray}
\widetilde{G}(\mathbf{r}^{\prime \prime },\mathbf{r}^{\prime }) &=&\frac{a}{%
r^{\prime \prime }}\underset{j,\kappa }{\sum }\frac{\Gamma \left( 1+\lambda +%
\widetilde{\epsilon }-\sqrt{\widetilde{\epsilon }^{2}-\frac{\omega ^{2}}{q}}%
\right) \Gamma \left( 1+\lambda +\widetilde{\epsilon }+\sqrt{\widetilde{%
\epsilon }^{2}-\frac{\omega ^{2}}{q}}\right) }{\Gamma \left( 2\widetilde{%
\epsilon }+1\right) \Gamma (2\lambda +2)}  \notag \\
&&U(r^{\prime \prime })\left\{ \left( \mu -\frac{\kappa E}{\gamma }+\frac{%
V_{0}\widetilde{\beta }}{e^{r^{\prime }/a}-q}-\frac{aV_{0}\widetilde{\beta }%
}{qr^{\prime }}\right) \frac{U(r^{\prime })}{r^{\prime }}\Omega _{\kappa
,\kappa }^{j}\left( \theta ^{\prime \prime },\phi ^{\prime \prime },\theta
^{\prime },\phi ^{\prime }\right) \mathbf{\beta }^{2}\right.  \notag \\
&&\left. -\widetilde{\beta }\left( \frac{d}{dr^{\prime }}+\frac{1+\widetilde{%
\beta }\gamma }{r^{\prime }}-\frac{V_{0}\widetilde{\beta }E}{\alpha q\gamma }%
\right) \frac{1}{r^{\prime }}U(r^{\prime })\Omega _{\kappa ,-\kappa
}^{j}\left( \theta ^{\prime \prime },\phi ^{\prime \prime },\theta ^{\prime
},\phi ^{\prime }\right) \mathbf{\alpha }_{1}\mathbf{\alpha }_{2}\mathbf{%
\alpha }_{3}\right\} ,  \notag \\
&&  \label{a49}
\end{eqnarray}%
where 
\begin{eqnarray}
U(r^{\prime }) &=&\left( qe^{-r^{\prime }/a}\right) ^{\widetilde{\epsilon }%
}\left( 1-qe^{-r^{\prime }/a}\right) ^{\lambda +1}  \notag \\
&&\times \text{ }_{2}F_{1}\left( 1+\lambda +\widetilde{\epsilon }-\sqrt{%
\widetilde{\epsilon }^{2}-\frac{\omega ^{2}}{q}},1+\lambda +\widetilde{%
\epsilon }+\sqrt{\widetilde{\epsilon }^{2}-\frac{\omega ^{2}}{q}},2\lambda
+2;1-qe^{-r^{\prime }/a}\right)  \notag \\
&&  \label{a50}
\end{eqnarray}%
and%
\begin{eqnarray}
U(r^{\prime \prime }) &=&\left( qe^{-r^{\prime \prime }/a}\right) ^{%
\widetilde{\epsilon }}\left( 1-qe^{-r^{\prime \prime }/a}\right) ^{\lambda
+1}  \notag \\
&&\times \text{ }_{2}F_{1}\left( 1+\lambda +\widetilde{\epsilon }-\sqrt{%
\widetilde{\epsilon }^{2}-\frac{\omega ^{2}}{q}},1+\lambda +\widetilde{%
\epsilon }+\sqrt{\widetilde{\epsilon }^{2}-\frac{\omega ^{2}}{q}},2%
\widetilde{\epsilon }+1;qe^{-r^{\prime \prime }/a}\right) .  \notag \\
&&  \label{a51}
\end{eqnarray}

The poles of the Green's function (\ref{a49}), coming from the first $\Gamma
-$function in the numerator, are%
\begin{equation}
1+\lambda +\widetilde{\epsilon }-\sqrt{\widetilde{\epsilon }^{2}-\frac{%
\omega ^{2}}{q}}=-n_{r},\text{ \ \ \ \ \ }n_{r}=0,1,2,...\text{ }.
\label{a52}
\end{equation}%
They determine the bound-state energies. Converting (\ref{a52}) into energy
by using (\ref{a34a}) and (\ref{a34b}) yields%
\begin{eqnarray}
\left( E_{n_{r},\kappa }-\frac{V_{0}}{2q}\right) ^{2} &=&\frac{\left(
n_{r}+\lambda +1\right) ^{2}}{\left( n_{r}+\lambda +1\right) ^{2}+\left( 
\frac{aV_{0}}{q}\right) ^{2}}\left[ \mu ^{2}+\frac{1}{12a^{2}}\kappa \left(
\kappa -\widetilde{\beta }\right) \right] -\frac{1}{4a^{2}}\left(
n_{r}+\lambda +1\right) ^{2}  \notag \\
&&  \label{a53}
\end{eqnarray}%
for $q\geq 1$ and where $\lambda $ has been given by (\ref{a16}).

\section{Particular cases}

\subsection{First case: standard Hulth\'{e}n potential}

By setting $q=1$ in the expression (\ref{a1}), we obtain the standard Hulth%
\'{e}n potential%
\begin{equation}
V_{1}(r)=-\frac{V_{0}}{e^{r/a}-1}.  \label{a54}
\end{equation}%
The parameters (\ref{a13a}) and (\ref{a34b}) can thus be written

\begin{equation}
\gamma =\pm \left[ \kappa ^{2}-\left( aV_{0}\right) ^{2}\right] ^{\frac{1}{2}%
},  \label{a55}
\end{equation}%
and

\begin{equation}
\omega =a\sqrt{V_{0}\left( V_{0}-2E\right) }.  \label{a56}
\end{equation}%
The Green's function satisfying the linear Dirac equation can be deduced
from (\ref{a49}):%
\begin{eqnarray}
\widetilde{G}(\mathbf{r}^{\prime \prime },\mathbf{r}^{\prime }) &=&\frac{a}{%
r^{\prime \prime }}\underset{j,\kappa }{\sum }\frac{\Gamma \left( 1+\lambda +%
\widetilde{\epsilon }-\sqrt{\widetilde{\epsilon }^{2}-\omega ^{2}}\right)
\Gamma \left( 1+\lambda +\widetilde{\epsilon }+\sqrt{\widetilde{\epsilon }%
^{2}-\omega ^{2}}\right) }{\Gamma \left( 2\widetilde{\epsilon }+1\right)
\Gamma (2\lambda +2)}  \notag \\
&&U(r^{\prime \prime })\left\{ \left( \mu -\frac{\kappa E}{\gamma }+\frac{%
V_{0}\widetilde{\beta }}{e^{r^{\prime }/a}-1}-\frac{aV_{0}\widetilde{\beta }%
}{r^{\prime }}\right) \frac{U(r^{\prime })}{r^{\prime }}\Omega _{\kappa
,\kappa }^{j}\left( \theta ^{\prime \prime },\phi ^{\prime \prime },\theta
^{\prime },\phi ^{\prime }\right) \mathbf{\beta }^{2}\right.  \notag \\
&&\left. -\widetilde{\beta }\left( \frac{d}{dr^{\prime }}+\frac{1+\widetilde{%
\beta }\gamma }{r^{\prime }}-\frac{aV_{0}\widetilde{\beta }E}{\gamma }%
\right) \frac{1}{r^{\prime }}U(r^{\prime })\Omega _{\kappa ,-\kappa
}^{j}\left( \theta ^{\prime \prime },\phi ^{\prime \prime },\theta ^{\prime
},\phi ^{\prime }\right) \mathbf{\alpha }_{1}\mathbf{\alpha }_{2}\mathbf{%
\alpha }_{3}\right\} ,  \notag \\
&&  \label{a57}
\end{eqnarray}

with

\begin{eqnarray}
U(r^{\prime }) &=&\left( e^{-r^{\prime }/a}\right) ^{\widetilde{\epsilon }%
}\left( 1-e^{-r^{\prime }/a}\right) ^{\lambda +1}  \notag \\
&&\times \text{ }_{2}F_{1}\left( 1+\lambda +\widetilde{\epsilon }-\sqrt{%
\widetilde{\epsilon }^{2}-\omega ^{2}},1+\lambda +\widetilde{\epsilon }+%
\sqrt{\widetilde{\epsilon }^{2}-\omega ^{2}},2\lambda +2;1-e^{-r^{\prime
}/a}\right)  \notag \\
&&  \label{a58}
\end{eqnarray}%
and%
\begin{eqnarray}
U(r^{\prime \prime }) &=&\left( e^{-r^{\prime \prime }/a}\right) ^{%
\widetilde{\epsilon }}\left( 1-e^{-r^{\prime \prime }/a}\right) ^{\lambda +1}
\notag \\
&&\times \text{ }_{2}F_{1}\left( 1+\lambda +\widetilde{\epsilon }-\sqrt{%
\widetilde{\epsilon }^{2}-\omega ^{2}},1+\lambda +\widetilde{\epsilon }+%
\sqrt{\widetilde{\epsilon }^{2}-\omega ^{2}},2\widetilde{\epsilon }%
+1;e^{-r^{\prime \prime }/a}\right) .  \notag \\
&&  \label{a59}
\end{eqnarray}

The energy spectrum is then found from (\ref{a53}) to be

\begin{eqnarray}
\!\!\!\left( E_{n_{r},\kappa }-\frac{V_{0}}{2}\right) ^{2} &=&\frac{\left(
n_{r}+\lambda +1\right) ^{2}}{\left( n_{r}+\lambda +1\right) ^{2}+\left(
aV_{0}\right) ^{2}}\left[ \mu ^{2}+\frac{\kappa \left( \kappa -\widetilde{%
\beta }\right) }{12a^{2}}\right] -\frac{\left( n_{r}+\lambda +1\right) ^{2}}{%
4a^{2}}.  \notag \\
&&  \label{a60}
\end{eqnarray}%
This result coincides with that obtained through the resolution of the Dirac
equation \cite{Haouat}, if we adopt the same approximation to the
centrifugal term which consists in missing out the quantity $\frac{\kappa
\left( \kappa -\widetilde{\beta }\right) }{12a^{2}}.$

\subsection{Second case: Coulomb potential}

When $q=1$ and $a$ $\rightarrow \infty $, the deformed Hulth\'{e}n potential
given by Eq. (\ref{a1}) turns to the potential of a hydrogen-like ion%
\begin{equation}
V_{c}(r)=-\frac{Ze^{2}}{r},  \label{a61}
\end{equation}%
where we have put $V_{0}=Ze^{2}/a$ with $Ze$ the charge of the nucleus.

In this case, it can be seen from Eqs. (\ref{a34a}) and (\ref{a34b}) that

\bigskip

\begin{equation}
\left\{ 
\begin{array}{c}
\widetilde{\epsilon }\underset{a\rightarrow \infty }{\simeq }a\sqrt{\mu
^{2}-E^{2}} \\ 
1+\lambda +\widetilde{\epsilon }-\sqrt{\widetilde{\epsilon }^{2}-\omega ^{2}}%
\underset{a\rightarrow \infty }{\simeq }1+\lambda _{0}-\frac{Ze^{2}E}{\sqrt{%
\mu ^{2}-E^{2}}} \\ 
1+\lambda +\widetilde{\epsilon }+\sqrt{\widetilde{\epsilon }^{2}-\omega ^{2}}%
\underset{a\rightarrow \infty }{\simeq }1+\lambda _{0}+\frac{Ze^{2}E}{\sqrt{%
\mu ^{2}-E^{2}}}+2a\sqrt{\mu ^{2}-E^{2}}\underset{a\rightarrow \infty }{%
\rightarrow }\infty .%
\end{array}%
\right.  \label{a62}
\end{equation}%
On the other hand, by using the Gauss transformation formula (see Ref. \cite%
{Gradshtein}, p. 1043, Eq. (9.131.2))

\begin{eqnarray}
_{2}F_{1}(a,b,c;z) &=&\frac{\Gamma (c)\Gamma (c-a-b)}{\Gamma (c-a)\Gamma
(c-b)}\text{ }_{2}F_{1}(a,b,a+b-c+1;1-z)  \notag \\
&&+\frac{\Gamma (c)\Gamma (a+b-c)}{\Gamma (a)\Gamma (b)}(1-z)^{c-a-b}\text{ }%
_{2}F_{1}(c-a,c-b,c-a-b+1;1-z),  \notag \\
&&  \label{a63}
\end{eqnarray}%
as well as the formula \cite{Landau}

\begin{equation}
\underset{\beta \rightarrow \infty }{\lim }\text{ }_{2}F_{1}(\alpha ,\beta
,\gamma ;\frac{z}{\beta })=\text{ }_{1}F_{1}(\alpha ,\gamma ;z)  \label{a64}
\end{equation}%
together with the connection between the confluent hypergeometric functions
and the standard Whittaker functions $M_{a,b}(z)$ and $W_{a,b}(z)$ (see Ref. 
\cite{Gradshtein}, p. 1059, Eqs. (9.220.2) and (9.220.4))%
\begin{equation}
\left\{ 
\begin{array}{c}
M_{a,b}(z)=z^{b+\frac{1}{2}}e^{-\frac{z}{2}}\text{ }_{1}F_{1}(b-a+\frac{1}{2}%
,2b+1;z), \\ 
W_{a,b}(z)=\frac{\Gamma (-2b)}{\Gamma (\frac{1}{2}-b-a)}M_{a,b}(z)+\frac{%
\Gamma (2b)}{\Gamma (\frac{1}{2}+b-a)}M_{a,-b}(z),%
\end{array}%
\right.  \label{a65}
\end{equation}%
it is easy to show that, in the limit $a\rightarrow \infty $, the closed
form (\ref{a45}) of the Green's function of the second order Dirac equation,
for $q=1$, becomes%
\begin{eqnarray}
\underset{a\rightarrow \infty }{\lim }\widetilde{g}(\mathbf{r}^{\prime
\prime },\mathbf{r}^{\prime }) &=&\widetilde{g}_{0}(\mathbf{r}^{\prime
\prime },\mathbf{r}^{\prime })  \notag \\
&=&\frac{1}{2r^{\prime \prime }r^{\prime }}\underset{j,\kappa }{\sum }\frac{%
\Gamma \left( 1+\lambda _{0}-\frac{Ze^{2}E}{\widetilde{\omega }}\right) }{%
\widetilde{\omega }\Gamma (2\lambda _{0}+2)}M_{\frac{Ze^{2}E}{\widetilde{%
\omega }},\lambda _{0}+\frac{1}{2}}(2\widetilde{\omega }r^{\prime })  \notag
\\
&&\times W_{\frac{Ze^{2}E}{\widetilde{\omega }},\lambda _{0}+\frac{1}{2}}(2%
\widetilde{\omega }r^{\prime \prime })\Omega _{\kappa ,\kappa }^{j}\left(
\theta ^{\prime \prime },\phi ^{\prime \prime },\theta ^{\prime },\phi
^{\prime }\right) \mathbf{\beta }^{2};\text{ \ }r^{\prime \prime }>r^{\prime
},  \notag \\
&&  \label{a66}
\end{eqnarray}%
where $\widetilde{\omega }=\sqrt{\mu ^{2}-E^{2}},$and $\lambda
_{0}=\left\vert \gamma _{0}\right\vert -\frac{1}{2}\left( sign\gamma
_{0}-1\right) ,$ with $\gamma _{0}=\pm \sqrt{\kappa ^{2}-Z^{2}e^{4}}$.

With the Coulomb potential, the operator (\ref{a46}) reduces to%
\begin{equation}
\mathbf{M}_{s}=i\mathbf{\beta }\alpha _{r}\left[ \frac{d}{dr}+\frac{1-\gamma
_{0}\mathbf{\beta }}{r}+\frac{Ze^{2}E}{\gamma _{0}}\mathbf{\beta }\right] +%
\frac{\kappa E}{\gamma _{0}}\mathbf{\beta ,}.  \label{a67}
\end{equation}%
and the Whittaker function $M_{\frac{Ze^{2}E}{\widetilde{\omega }},\lambda
_{0}\left( \gamma _{0}\right) +\frac{1}{2}}(2\widetilde{\omega }r)$ in (\ref%
{a66}) verifies the recurrence relations \cite{Biedenharn,Kayed,Wong}

\begin{eqnarray}
D_{\pm }\left[ \frac{1}{r}M_{\frac{Ze^{2}E}{\widetilde{\omega }},\lambda
_{0}\left( \pm \gamma _{0}\right) +\frac{1}{2}}(2\widetilde{\omega }r)\right]
&=&\pm i\text{ }sgn\gamma _{0}\left[ \mu ^{2}-\left( \frac{\kappa E}{\gamma
_{0}}\right) ^{2}\right] ^{\frac{1}{2}}\frac{1}{r}M_{\frac{Ze^{2}E}{%
\widetilde{\omega }},\lambda _{0}\left( \mp \gamma _{0}\right) +\frac{1}{2}%
}(2\widetilde{\omega }r),  \notag \\
&&  \label{a68}
\end{eqnarray}%
where $D_{\pm }=\frac{d}{dr}+\frac{1\pm \gamma _{0}}{r}\mp \frac{Ze^{2}E}{%
\gamma _{0}}.$

In this case, the Green's function associated with the Dirac-Coulomb problem
is%
\begin{eqnarray}
\widetilde{G}_{0}(\mathbf{r}^{\prime \prime },\mathbf{r}^{\prime })
&=&\left( \mu +\mathbf{M}_{s}(r^{\prime })\right) \widetilde{g}_{0}(\mathbf{r%
}^{\prime \prime },\mathbf{r}^{\prime })  \notag \\
&=&\frac{1}{2r^{\prime \prime }r^{\prime }}\underset{j,\kappa }{\sum }\frac{%
\Gamma \left( 1+\lambda _{0}-\frac{Ze^{2}E}{\widetilde{\omega }}\right) }{%
\widetilde{\omega }\Gamma (2\lambda _{0}+2)}W_{\frac{Ze^{2}E}{\widetilde{%
\omega }},\lambda _{0}+\frac{1}{2}}(2\widetilde{\omega }r^{\prime \prime }) 
\notag \\
&&\times \left\{ \left( \mu -\frac{\kappa E}{\gamma _{0}}\right) M_{\frac{%
Ze^{2}E}{\widetilde{\omega }},\lambda _{0}+\frac{1}{2}}(2\widetilde{\omega }%
r^{\prime })\Omega _{\kappa ,\kappa }^{j}\left( \theta ^{\prime \prime
},\phi ^{\prime \prime },\theta ^{\prime },\phi ^{\prime }\right) \mathbf{%
\beta }^{2}\right.  \notag \\
&&\left. -i\widetilde{\omega }\text{ }sgn\gamma _{0}\text{ }M_{\frac{Ze^{2}E%
}{\widetilde{\omega }},\widetilde{\lambda }_{0}+\frac{1}{2}}(2\widetilde{%
\omega }r^{\prime })\Omega _{\kappa ,-\kappa }^{j}\left( \theta ^{\prime
\prime },\phi ^{\prime \prime },\theta ^{\prime },\phi ^{\prime }\right) 
\mathbf{\alpha }_{1}\mathbf{\alpha }_{2}\mathbf{\alpha }_{3}\right\} , 
\notag \\
&&  \label{a69}
\end{eqnarray}%
with $\lambda _{0}=\lambda _{0}\left( \gamma _{0}\right) $ and $\widetilde{%
\lambda }_{0}=\lambda _{0}\left( -\gamma _{0}\right) .$

The discrete energy spectrum is then%
\begin{equation}
E_{n_{r},\kappa }=\mu \left[ 1+Z^{2}e^{4}\left( n_{r}+\lambda _{0}+1\right)
^{-2}\right] ^{-\frac{1}{2}}.  \label{a70}
\end{equation}%
These results coincide with those obtained by path integration \cite{Kayed}.

\section{Conclusion}

In this contribution, we have shown that the problem of a relativistic spin-$%
\frac{1}{2}$ system in the presence of a deformed Hulth\'{e}n potential can
be solved for arbitrary $\kappa -$states by using a proper approximation to $%
\frac{1}{r^{2}}$. It is worth mentioning that the problem of a Dirac
particle moving in the standard Hulth\'{e}n potential cannot be solved
exactly even for the $s-$wave because of the centrifugal \ potential term
contained in the two radial wave equations (see for example Eqs. (47) in
Ref. \cite{Haouat}) contrary to what was stated by Guo et al. \cite{Guo} and
Alhaidari \cite{Alhaidari}. These authors had performed several incorrect
manipulations by applying a constraint to get rid of the centrifugal term
depending on the eigenvalues of the Martin-Glauber oprerator \cite%
{Martin,Wong}. Consequently, their energy spectra are unsatisfactory.
Furthermore, one can observe that the $s-$wave relativistic energy spectrum
for the Coulomb potential cannot be found from that obtained through
Alhaidari's approach. For the standard Hulth\'{e}n potential $(q=1)$ and the
Coulomb potential $(q=1$ and $a\rightarrow \infty )$, our results concerning
the energy spectra are identical with that obtained by solving Dirac's
equation \cite{Haouat} (see also Refs. \cite{Kayed, Schiff} \ for the
Coulomb potential).

\end{document}